\documentclass[twocolumn,showpacs,prb]{revtex4-1}
\usepackage{bm}
\usepackage{graphicx}
\usepackage{epsfig}
\usepackage{amsmath} 
\usepackage{color}
\usepackage{amssymb}
\usepackage{array}
\usepackage[colorlinks = true]{hyperref}

\newcommand{\eps}{\varepsilon}

\newcommand{\ket}[1]{\left| #1 \right\rangle}
\newcommand{\X}{\mathrm X}

\begin{document}

\title{Magneto-spectroscopy of excited states in charge-tunable GaAs/AlGaAs [111] quantum dots}

\author{M.\,V.\,Durnev$^1$}
\author{M. Vidal$^2$}
\author{L. Bouet$^2$}
\author{T. Amand$^2$}
\author{M.\,M.\,Glazov$^1$}
\author{E.\,L.\,Ivchenko$^1$}
\author{P. Zhou$^2$}
\author{G. Wang$^2$}
\author{T. Mano$^3$}
\author{N. Ha$^3$}
\author{T. Kuroda$^3$}
\author{X. Marie$^2$}
\author{K. Sakoda$^3$}
\author{B. Urbaszek$^2$}

\affiliation{$^1$Ioffe Institute, 194021 St.\,Petersburg, Russia}
\affiliation{$^2$Universit\'e de Toulouse, INSA-CNRS-UPS, LPCNO, 135 Avenue Rangueil, 31077 Toulouse, France}
\affiliation{$^3$National Institute for Material Science, Namiki 1-1, Tsukuba 305-0044, Japan}

\begin{abstract}
We present a combined experimental and theoretical study of highly charged and excited electron-hole complexes in strain-free (111) GaAs/AlGaAs quantum dots  grown by droplet epitaxy. We address the complexes with one of the charge carriers residing in the excited state, namely, the ``hot'' trions X$^{-*}$ and X$^{+*}$, and the doubly negatively charged exciton X$^{2-}$. Our magneto-photoluminescence experiments performed on single quantum dots in the Faraday geometry uncover characteristic emission patterns for each excited electron-hole complex, which are very different from the photoluminescence spectra observed in (001)-grown quantum dots. We present a detailed theory of the fine structure and magneto-photoluminescence spectra of X$^{-*}$, X$^{+*}$ and X$^{2-}$ complexes, governed by the interplay between the electron-hole Coulomb exchange interaction and the heavy-hole mixing, characteristic for these quantum dots with a trigonal symmetry. Comparison between experiment and theory of the magneto-photoluminescence allows for precise charge state identification, as well as extraction of electron-hole exchange interaction constants and $g$-factors for the charge carriers occupying excited states.
\end{abstract}

\maketitle 

%************************************************
\section{Introduction}\label{sec:intro}
%************************************************
Semiconductor quantum dots (QDs) have been investigated for more than two decades. Due to their nanometer size, QDs are true quantum emitters and single spin memories \cite{Nowak:2014a,Luxmoore:2013a,Nilsson:2013b,Degreve:2012a,Poem2010} and ideally suited for fundamental studies of quantum phenomena on a single particle level and for experiments in quantum optics and information processing.\cite{Schulte:2015fk, Delteil2015} A major breakthrough for QD studies is the embedding of the QD in a charge tunable device.\cite{Warburton:2000a,Ware:2005a,Ediger:2007a,Jovanov:2011a,Sanada:2009a,Dufaker:2013a,VanHattem:2013a} In these structures highly charged complexes with one hole or electron in an excited state can be studied.\cite{Urbaszek:2003a,Karrai:2004a,Ware:2005a,Ediger:2007a,Chang:2009a,Mlinar:2009a,Warming:2009a} Voltage control of the QD charge state permits clearer identification of the transitions as compared to dots with residual doping. These devices also permit to separate various electron-hole complexes for studying them independently.
Most of the studies were done on QDs, grown on (001) GaAs in Stranski-Krastanov process, leading to strained dots with disk or lens shape, characterized by the C$_{2v}$ point symmetry. Droplet epitaxy technique allows a more flexible approach~\cite{Koguchi:1991a, Wang:2007a, Kumah:2009a, Graf:2014a, Cavigli:2012a} and, particularly, for the growth of strain free GaAs dots on (111) surface~\cite{Mano:2010a, Stock:2010a, Treu:2012a}.

The advantage of strain free (111)-grown droplet GaAs/AlGaAs QDs is their high C$_{3v}$ point symmetry. It results in vanishingly small fine structure splittings of the exciton radiative doublet,\cite{Singh2009,Mano:2010a,Karlsson2010} which is a necessary requirement for the emission of entangled photon pairs from the biexciton-exciton cascade.\cite{Kuroda:2013a,Juska:2013a}  In addition, the C$_{3v}$ symmetry with only three-fold rotation axis $z\parallel [111]$ leads to a coupling between the two valence heavy-hole states $|\pm 3/2\rangle$ in a magnetic field applied along the [111] direction, i.e. in the Faraday geometry. This  allows for the simultaneous detection of the initially ``dark'' and ``bright'' excitonic states in magneto-photoluminescence (magneto-PL) experiments.\cite{Sallen:2011a,Oberli:2012a,Durnev:2013a}

While neutral excitons, biexcitons and singly charged excitons, also known as trions, are well studied experimentally and theoretically, only few works report on the magneto-PL of highly charged complexes on C$_{2v}$ (001)-grown quantum dots.\cite{Warburton:2000a,Jovanov:2011a,Dufaker:2013a,VanHattem:2013a} For C$_{3v}$ (111)-grown dots only magneto-PL of neutral (X$^0$) and charged excitons (X$^{+}$ with two holes, X$^{-}$ with two electrons) as well as of biexciton (XX$^0$) was reported.\cite{Sallen:2011a,Oberli:2012a, Durnev:2013a} The charge tuning has been demonstrated for (111) QDs quite recently,\cite{Bouet:2014a} but the magneto-PL of excited and highly charged electron-hole complexes has not been studied in detail so far.

Here we report on experimental and theoretical studies of magneto-optical properties of some of the simplest charged complexes with charge carriers, residing in the excited states ($p$-shell in atom-like nomenclature) of (111)-grown quantum dots. We address the ``hot'' trions, X$^{-*}$ and X$^{+*}$ with one of two identical charge carriers occupying the first excited state as well as the doubly charged exciton complex, X$^{2-}$, with three electrons and one hole. We demonstrate that using comparison between the experiment and the theory, the magneto-PL spectroscopy allows for precise charge state identification. We find, that the rich structure of magneto-PL spectra of ``hot'' and multiple charged electron-hole complexes is a consequence of the heavy-hole mixing, present in our highly symmetric (111) dots.
Comparison between the experiment and developed theory allows us to extract the parameters governing the spin-dependent fine structure of electron-hole complexes, i.e. electron-hole exchange constants and components of the $g$-factor tensor for electrons and holes occupying excited states of the QD. 

This paper is organized as follows: We first introduce the sample and the magneto-PL setup in Sec.~\ref{sec:expdetails}. The main points of our theory are outlined in Sec.~\ref{sec:theory_general}. The key features observed in magneto-PL and the detailed theory for X$^{-*}$, X$^{+*}$ and X$^{2-}$ complexes are presented in Sec.~\ref{sec:theory_exp}. Section~\ref{sec:discuss} contains the discussion of our results and the paper is summarized in Sec.~\ref{sec:concl}.

%************************************************
\section{Experimental details}\label{sec:expdetails}
\begin{figure*}[htpb]
\includegraphics[width=0.7\linewidth]{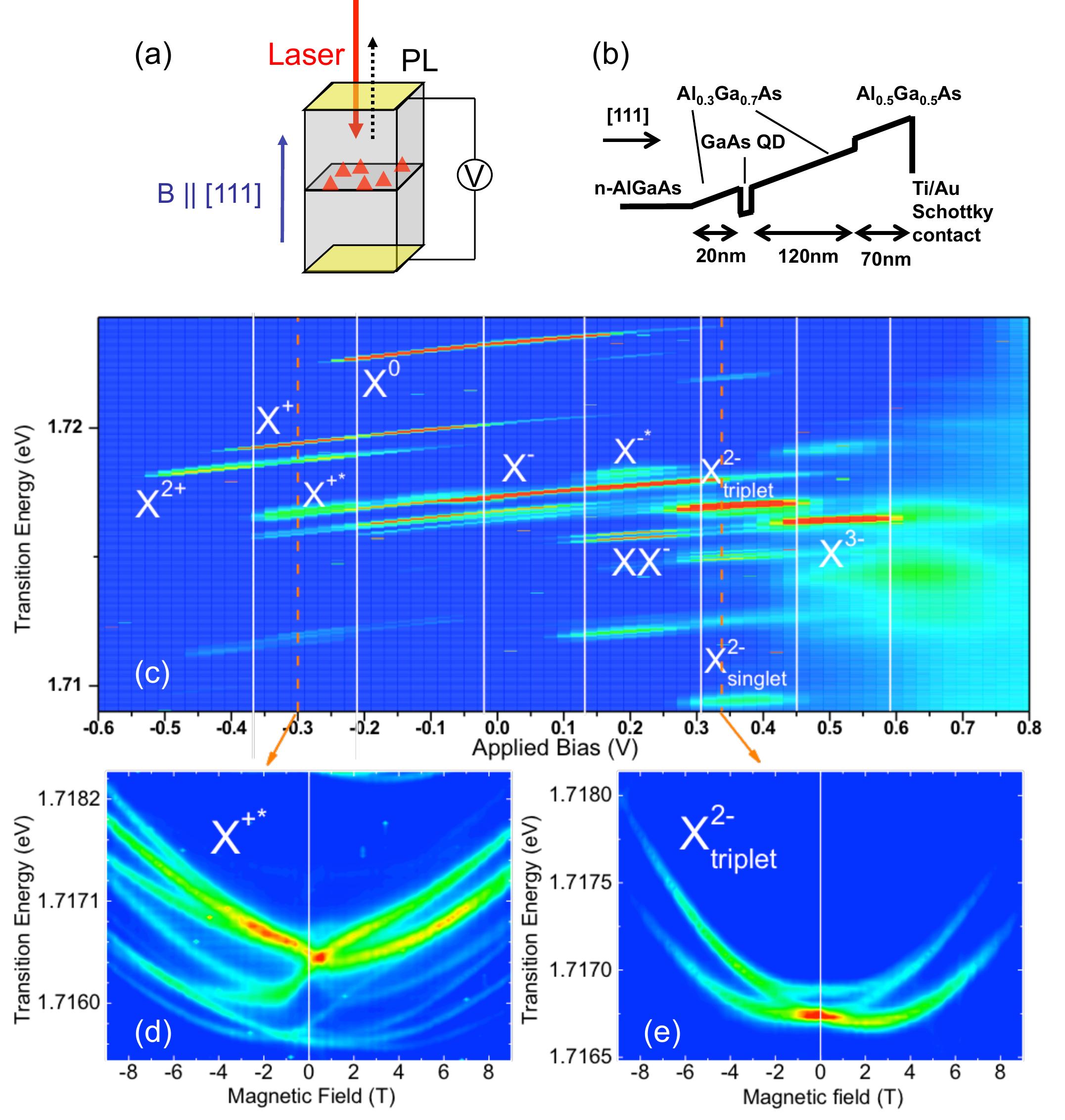}
\caption{\label{fig:fig1} PL spectra of QD P in a charge tunable device. (a) The schematic structure of the sample and experimental geometry: The dots are located between the bottom gate (doped layer) and the semitransparent top gate. (b) The schematic structure of the charge tunable device with the conduction band profile. (c) The contour plot of the QD P PL spectrum as a function of bias at zero magnetic field. The main transitions are labelled as in Ref.~\onlinecite{Bouet:2014a}. The other identified positively and negatively charged complexes are also labelled (see the text for details). Here blue corresponds to $<100$ counts, red to $>3000$ counts.  Panels (d) and (e) show two typical magneto-PL patterns at fixed bias as a function of the applied magnetic field along [111]. In panel (d)  blue corresponds to $<100$ counts, red to $>12800$ counts, in panel (e)  blue corresponds to $<100$ counts, red to $>40000$ counts.}
\end{figure*}

%************************************************
The sample used in this study was grown by droplet epitaxy using a standard molecular beam epitaxy system.\cite{Mano:2010a,Sallen:2011a} 
Starting from the $n^+$-GaAs(111)A substrate, the sample consists of 50-nm $n$-GaAs (Si: $1\times10^{18}$~cm$^{-3}$), 100-nm $n$-Al$_{0.3}$Ga$_{0.7}$As (Si: $1\times10^{18}$~cm$^{-3}$), 20-nm Al$_{0.3}$Ga$_{0.7}$As tunnel barrier, GaAs QDs, 120-nm Al$_{0.3}$Ga$_{0.7}$As, 70-nm Al$_{0.5}$Ga$_{0.5}$As, and 10-nm GaAs cap. The sample structure, experimental geometry and conduction band diagram are schematically illustrated in Fig.~\ref{fig:fig1}(a) and (b).

The morphology analysis using AFM reveals symmetric dots with a typical height of $\simeq$ 2\dots3~nm and a radius of $\simeq$ 15~nm.\cite{Durnev:2013a} In this model system dots are truly isolated as they are not connected by a 2D wetting layer,
\cite{Mano:2010a,Sallen:2014a} contrary to Stranski-Krastanov dots and QDs formed at quantum well interface fluctuations.\cite{Bracker:2005a} A semitransparent Ti/Au layer with a nominal thickness of 6 nm serves as a Schottky top gate for charge tuneability of the device, see Fig.~\ref{fig:fig1}(b). Application of a bias voltage allows the controlled charging of the QDs from $-3|e|$ to $+2|e|$.\cite{Bouet:2014a} 

Single dot PL at 4K is recorded with a home-build confocal microscope with a detection spot diameter of
$\simeq 1$~$\mu$m.~\cite{Durnev:2013a,Sallen:2014a} The detected PL signal is dispersed by a spectrometer and detected by a Si-CCD camera (spectral precision of 1~$\mu$eV and spectral resolution around 50~$\mu$eV). Optical excitation is achieved by pumping the AlGaAs barrier with a HeNe laser at 1.96~eV. Laser polarization control and PL polarization analysis is performed with Glan-Taylor polarisers and liquid crystal waveplates. Magnetic fields up to 9~T can be applied parallel to the growth axis [111] (Faraday geometry), that is also the angular momentum quantization axis and the light propagation axis. In what follows we denote different QDs in the sample by capital Latin letters, A, B, C, etc. The experimental spectra are presented for several characteristic quantum dots, namely, C, F, H, K, N, P, U, X.

Figure~\ref{fig:fig1} shows the results of charge tuning in our sample for one of the QDs, i.e., QD P, as we present its PL contour plot as a function of applied voltage, see Fig.~\ref{fig:fig1}(c).
We observe the main exciton complexes X$^+$, X$^0$, X$^-$, X$^{2-}$ and X$^{3-}$ as well as two other groups of transitions appearing at the same bias as X$^+$ and X$^-$. Based on comparison with the theory presented in Sec.~\ref{sec:theory_exp}, we identify these two complexes as the ``hot'' trion states X$^{+*}$ and X$^{-*}$ with one charge carrier (hole or electron, respectively), residing in an excited shell. 
We have performed magneto-PL for all observed complexes. The intriguing magneto-PL of X$^+$, X$^0$ and X$^-$ has been analyzed previously revealing heavy-hole mixing~\cite{Sallen:2011a,Durnev:2013a} in a longitudinal magnetic field. Here we concentrate on the very unusual magneto-PL of transitions of ``hot'' and highly charged complexes, such as the X$^{+*}$ shown in Fig.~\ref{fig:fig1}(d) and X$^{2-}$ shown in Fig.~\ref{fig:fig1}(e). Our theory, introduced in the next two sections, aims to reproduce the evolution of the main transitions observed as a function of applied magnetic field.

%\begin{table}
%\caption{\label{tab:table2} Parameters used in theoretical modeling of complexes presented in Fig.~\ref{fig:fig2}: $p$-states $g$-factors ($g_e^{(p)}$,$g_{h1}^{(p)}$,$g_{h2}^{(p)}$), $s$-$p$ exchange constants ($\delta_0^{sp}$, $\tilde{\delta}_0^{sp}$), diamagnetic shifts ($\alpha_d$) and line broadenings ($\gamma$).}
%\begin{ruledtabular}
%\begin{tabular}{cccc}
% & X$^{-*}$ (QDC) & X$^{+*}$ (QDK) & X$^{2-}$ (QDP) \\
% \hline
%$g_e^{(p)}$ & 0.46 & -- & 0.32\\
%$g_{h1}^{(p)}$ & -- & 2.5 & --\\ 
%$g_{h2}^{(p)}$ & -- & 0.38 & --\\
%$\delta_0^{sp}$ ($\mu$eV) & \MD{10 (?)} & -- & \MD{60} \\
%$\delta_1^{sp}$ ($\mu$eV) & -- & -- & \MD{60} \\
%$\tilde{\delta}_0^{sp}$ ($\mu$eV) & -- & \MD{150} & -- \\
%$\alpha_d$ ($\mu$eV/T$^2$) & 5 & 13 & 9 \\
%$\gamma$ ($\mu$eV) & 25 & 40 &  30\\
%\end{tabular}
%\end{ruledtabular}
%\end{table}

%\begin{figure*}[htpb]
%\includegraphics[width=\linewidth]{Figs/Fig2.pdf}
%\caption{\label{fig:fig2} Magneto-PL patterns of three recognized excited electron-hole complexes: negatively charged trion X$^{-*}$ (panels a and d), positively charged trion X$^{+*}$ (panels b and e), and doubly negatively charged trion X$^{2-}$ (panels c and f). The experimental spectra are shown in panels a-c for QDC, QDK and QDP, and the results of theoretical simulation -- in panels d-f. The insets in panels d-f schematically illustrate the structure of the complexes.
%}
%\end{figure*}

%************************************************
\section{Theory: general model}\label{sec:theory_general}
%************************************************
This section is aimed at the description of basic features of the theoretical model for the energy and PL-spectra of the hot and multiply charged electron-hole complexes in trigonal quantum dots in the presence of a longitudinal magnetic field $\bm B \parallel [111]$. The detailed theory as well as experimental spectra for each of the studied complexes, namely, ``hot'' charged trions X$^{-*}$ and X$^{+*}$, and a doubly negatively charged trion X$^{2-}$, are given in Sec.~\ref{sec:theory_exp}. 

We start with the description of quantum confinement of many particle electron-hole complexes in trigonal quantum dots and follow the model developed in Ref.~\onlinecite{Durnev:2013a} and applied to evaluate hole wave functions and $g$-factors.
In our model we assume strong confinement of electrons and holes inside a dot, which allows one to treat the Coulomb interaction as a small perturbation, which influences essentially (i) the spectral positions of emission of corresponding electron-hole complexes via its binding energy,\footnote{Binding energy evaluation is beyond the scope of the present paper.} and (ii) the fine spin structure of the electron-hole energy spectrum via electron-electron (hole-hole) and electron-hole exchange interaction being, along with the Zeeman splittings, the focus of our work.
This approach is valid for quantum dots with a size smaller than the exciton Bohr radius, which is typically the case in our experiment.\cite{Durnev:2013a} To model confinement in the dot we assume the separable form for a single-particle potential $V_{e,h}(\bm r_{e,h})=V^{(e,h)}_\perp(z_{e,h}) + V^{(e,h)}_\parallel (\bm \rho_{e,h})$ acting on the electrons, $e$, and holes, $h$,  respectively, $\bm r_{e,h}$ is the position vector of corresponding charge carrier. Here $V_\perp(z)$ and $V_\parallel (\bm \rho)$ describe carrier confinement in the $z$-direction and in the lateral plane of a dot, respectively, and we use cylindrical coordinates $\bm r = (\bm \rho, z)$, with the $z$-axis parallel to the growth axis [111]. The in-plane axes are chosen as $x\parallel [11\bar2]$ and $y\parallel [\bar 110]$. This separable form is a good approximation for our trigonal dots having shapes of flattened pyramids with its lateral size much larger than its height.\cite{Mano:2010a, Durnev:2013a}

The form of the confining potentials can be chosen following Ref.~\onlinecite{Durnev:2013a} in accordance with the trigonal point symmetry $C_{3v}$ of the QD. Particularly, the potential $V_\perp(z)$ has no specific parity with respect to the mirror reflection $z \to -z$ and can be taken in the form of a triangular well. The in-plane potential can be taken as a combination of an isotropic parabolic potential\cite{que92,hawrylak99,semina06} and a trigonal contribution proportional to $\cos{3\varphi}$ with $\varphi$ being the azimuthal angle of the position vector reckoned from $x$-axis.

The separable form of a confining potential allows one to factorize the wave function $\Psi(\bm r)$ of an electron (hole) localized in the quantum dot into a product of $z$- and $\bm \rho$-dependent parts
\begin{equation}
\label{psi:gen}
\Psi_{nl}(\bm r) = F_n(z)\psi_l(\bm \rho)\:.
\end{equation}
Here indices $n = 1,2,\dots$ and $l = 1,2,\dots$ are the quantum numbers characterizing confinement of the $z$- and in-plane motion, respectively. Since in our dots confinement in the growth direction is much stronger than in the lateral plane, the first excited states in the dot are the states corresponding to $n=1$, $l>1$.\cite{Durnev:2013a} The symmetry of the single-particle states underlies the fine structure of the electon-hole complexes energy spectrum and their magneto-optical properties. Therefore, it is instructive to classify $\Psi_{nl}$ according to the irreducible representations of $C_{3v}$ point symmetry group. Since any function $F_n(z)$ transforms according to the identity representation, $\Gamma_1$ (or $A_1$),\cite{koster63} the transformation properties of the wave function~\eqref{psi:gen} are governed by the symmetry of the in-plane envelope $\psi_l(\bm \rho)$. Due to the presence of the three-fold rotation axis, the in-plane envelopes can transform either according to $\Gamma_1$ or according to the two-fold degenerate representation $\Gamma_3$ ($E$). In atom-like nomenclature or shell model, widely used to classify the states in QDs,\cite{bayer98,hawrylak99,PhysRevB.75.205313} $\Gamma_1$ state can be attributed to the $s$-shell and the $\Gamma_3$ doublet to the two-fold degenerate $p$-shells. 

In what follows we consider ground $s$-type state ($l=1$) and two excited $p$-type states ($l=2,3$) of the in-plane motion. In order to simplify further treatment we assume that the base of a dot is slightly elongated in one direction, so that the degeneracy of the $p$-states is lifted.
%\footnote{\addMisha{Experiments indeed reveal small but finite anisotropic splitting of exciton radiative doublet confirming slight anisotropy of the dots.\cite{Sallen:2011a,Kuroda:2013a,Liu:2014a}}} 
 By analogy with elliptically shaped C$_{2v}$ dots, we refer to the corresponding states as $p_x$ and $p_y$, and in the following we will consider only one state with a lower energy, i.e., $p_x$. As we will show later, in Sec.~\ref{sec:discuss}, the magneto-PL spectra of X$^{-*}$ and X$^{+*}$ do not change when the orbital degeneracy of $p$-shell states is taken into account.
  
 The fine structure of the electron-hole complexes in the studied dots is governed by an interplay of the electron-hole exchange interaction, whose precise form is detailed below in Sec.~\ref{sec:theory_exp} where the particular complexes are identified, and Zeeman effects for individual charge carriers. For the conduction band electrons, whose Bloch functions transform according to the spinor representation $\Gamma_4$, the Zeeman effect in $\bm B \parallel [111]$ can be described as
 \begin{equation}
 \label{electron:Z}
 \mathcal H_B^e = \frac{1}{2} g_e^{(nl)} \mu_B B_z \sigma_z^{(e)},
 \end{equation}
 where $\mu_B$ is the Bohr magneton, $\sigma_z^{(e)}$ is the $z$-component Pauli matrix, and $g_e^{(nl)}$ (in what follows $g_e^{(s)}$ or $g_e^{(p)}$) is the electron $g$-factor in $nl$ orbital state. 
% The Zeeman effect for holes is more complex and the magneto-induced mixing of heavy holes (this doublet transforms according to the reducible $\Gamma_5 + \Gamma_6$ representation of the $C_{3v}$ point group) with opposite projections of angular momenta onto the growth axis $\pm 3/2$ is of particular importance. 
The doubly degenerate hole state transforms according to the reducible $\Gamma_5 + \Gamma_6$ representation of the C$_{3v}$ point group, and the magneto-induced mixing of heavy holes with opposite projections $\pm 3/2$ of angular momentum onto the growth axis is of top importance. 
The latter effect is a specific feature of the trigonal C$_{3v}$ symmetry, and was uncovered experimentally in Ref.~\onlinecite{Sallen:2011a} followed by the microscopic theory presented in Ref.~\onlinecite{Durnev:2013a}. As a consequence of the heavy-hole mixing, the effective 
Zeeman Hamiltonian for heavy holes ($n=1$) in the field $\bm B \parallel [111]$ has the form~\cite{Sallen:2011a,Durnev:2013a}
\begin{multline}
\label{eq:hh_mixing}
\mathcal H_B^{h} = \frac{1}{2} \mu_B B_z (g_{h1}^{(l)} \sigma_z^{(h)} + g_{h2}^{(l)} \sigma_x^{(h)}) = \\
= \frac12 \mu_B B_z 
\left(
\begin{array}{cc}
g_{h1}^{(l)} & g_{h2}^{(l)} \\
g_{h2}^{(l)} & -g_{h1}^{(l)} \\
\end{array}
\right)\:,
\end{multline}
so that the effective $g$-factor tensor of the heavy hole contains both the diagonal ($g_{h1}^{(l)}$) and off-diagonal ($g_{h2}^{(l)}$) components (in what follows $l = s, p$). Here $\bm \sigma^{(h)} = (\sigma_x^{(h)},\sigma_y^{(h)},\sigma_z^{(h)})$ is the pseudovector composed of Pauli matrices acting in the space of $\pm 3/2$ Bloch functions of heavy holes. As it was established in previous studies on neutral excitons X$^0$ and ground trion states X$^-$ and X$^+$, the values of $g_{h1}$ and $g_{h2}$ for the $s$-hole are of the same order of magnitude.\cite{Sallen:2011a, Durnev:2013a}
 
To calculate the PL-spectra of the studied electron-hole complexes we apply the model similar to the model of Refs.~\onlinecite{PhysRevB.75.205313,Glazov:2010a}. We consider a quantum dot that is non-resonantly excited with a continuous-wave laser, so that the steady-state populations of the spin states are formed. The PL spectra in $\sigma^+$ and $\sigma^-$ polarizations are given by the Fermi golden rule:
\begin{equation}
\label{eq:PL}
I_{\pm} (\omega)= A \sum \limits_{if} \left| M_{fi}^\pm \right|^2 n(E_i) \Delta(E_i - E_f + \delta E_{\rm dia} - \hbar \omega)\:.
\end{equation}
Here $| i \rangle$ is the initial state of the complex, $| f \rangle$ is the final state  of the QD after optical recombination, $M_{fi}^\pm$ is the matrix element of an optical transition, defined by selection rules and wave functions of $i$ and $f$ states, $E_{i,f}$ are the energies of the initial and final states, $\hbar \omega$ is the energy of an emitted photon, and $A$ is a constant. Note, that we consider here only the radiative recombination processes involving an electron and a hole in the ground shells, i.e., $s$-shell electron and $s$-shell hole. To include the diamagnetic shift of electron-hole complexes levels we introduce an energy shift $\delta E_{\rm dia} = \alpha_d B_z^2$ common for all lines of a given complex, where $\alpha_d$ is a parameter. Its evaluation as well as the evaluation of precise energy positions of the initial and final states are beyond the scope of the present paper.
While calculating the PL intensity we assume the initial levels $i$ to be equally populated, i.e. take for the occupation numbers $n(E_i) = 1$.
To model the broadening of PL lines we use the Lorentzian 
\[
\Delta(E) = \frac{1}{\pi} \frac{\gamma}{E^2+\gamma^2},
\]
with the broadening parameter $\gamma$. This parameter may, in general, be different for different initial and final states of a given complex, however in the model we use a single value of $\gamma$ for all summands in Eq.~\eqref{eq:PL}.

% \addMael{The g-factor of s-state for electron and hole and the fine structure constant used in the complexes modelisation was extracted experimentally, see Section \ref{sec:expdetails} for more details.}

%************************************************
\section{Identification of the complexes}\label{sec:theory_exp}
%************************************************

In this section we present detailed theoretical and experimental description of the three excited complexes, which are the most pronounced in the experimental PL spectra, namely, the ``hot'' charged trions X$^{-*}$ and X$^{+*}$, which involve electron or hole in the excited shell, and the doubly charged exciton X$^{2-}$. The results are presented in a form of comparison between the theory and experiment for three different, representative QDs. Such a comparative analysis allowed us to identify these three complexes in magneto-PL experiments for many QDs. Experimental PL-spectra and extracted fine structure parameters for an additional set of QDs are presented in Sec.~\ref{sec:discuss}.

%************************************************
\subsection{Hot negatively charged trion X$^{-*}$}\label{sec:Xminstar}
%************************************************

\begin{figure}[htpb]
\includegraphics[width=0.5\textwidth]{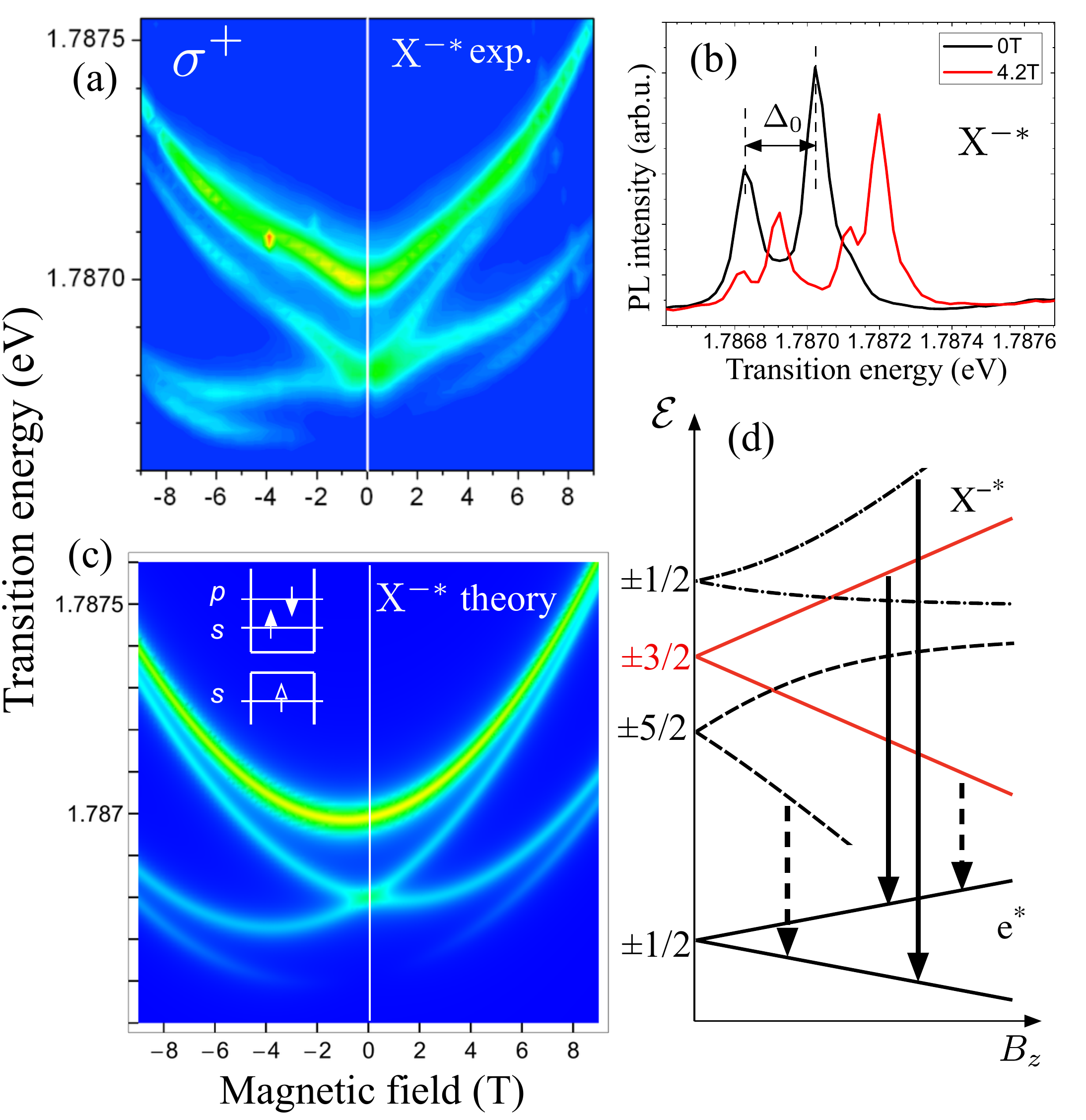}
\caption{\label{fig:Xminstar} Summary of theoretical and experimental results for magneto-PL of the X$^{-*}$ complex (QD C). (a) Measured PL pattern in $\sigma^+$ polarization and (b) measured PL spectra in $\sigma^+$ polarization for two values of magnetic field, (c) theoretical calculation of the PL pattern, (d) energy levels and schematic illustration of selection rules in $\sigma^+$ polarization.
The lower part of the panel (d) shows energy levels of the $p$-electron ($e^*$) which remains after X$^{-*}$ recombination; dashed arrows depict the transitions that become active due to heavy-hole mixing, i.e. $g_{h2}^{(s)} \neq 0$. The parameters used in calculations are $\Delta_0 = 190$~$\mu$eV, $g_e^{(p)} = 0.47$, $\gamma = 25$~$\mu$eV, $\alpha_d = 5$~$\mu$eV/T$^2$, where the value of $\Delta_0$ is directly taken from the spectrum shown in panel (b).
}
\end{figure}

Let us start our description with the ``hot'' negatively charged trion X$^{-*}$. This complex comprises a heavy hole in the $s$-state and two electrons, one of which occupies the ground, $s$-shell state and the other resides in the excited $p$-state. The schematic structure of the complex and calculated (see below for details) magneto-PL spectrum are depicted in Fig.~\ref{fig:Xminstar}(c). For comparison, the experimental magneto-PL spectrum of X$^{-*}$ measured in $\sigma^+$ polarization is presented in Fig.~\ref{fig:Xminstar}(a). The spectrum in $\sigma^-$ polarization is obtained from the one presented in Fig.~\ref{fig:Xminstar}(a) by changing the sign of the magnetic field. At zero magnetic field the X$^{-*}$ PL spectrum consists of two lines (a doublet) separated by 100\dots200~$\mu$eV with the higher energy transition having higher intensity. When a magnetic field $\bm B = (0,0,B_z)$ is applied, the lower energy line is split into two lines for each detected circular polarization, while an additional low intensity line appears at lower energy. 
%As a result, the X$^{-*}$ emission in a magnetic field consists of four lines, active in each, $\sigma^+$ and $\sigma^-$, polarization, see Fig.~\ref{fig:Xminstar}(b) for a typical emission spectrum found in the experiment. 
As a result, in each, $\sigma_+$ or $\sigma_-$ polarization, the X$^{-*}$ emission consists of four lines, see Fig.~\ref{fig:Xminstar}(b) for a typical emission spectrum found in the experiment, to give eight lines for the linearly polarized analyzer. 
As we will show later, the appearance of the low intensity (dark at $B_z = 0$) line is a characteristic feature of the underlying C$_{3v}$ symmetry of the dots, resulting from the heavy-hole mixing.

We now focus on theoretical description of X$^{-*}$. The spin states of X$^{-*}$ are defined by the $z$-component of the heavy-hole angular momentum ($J_z = \pm 3/2$), the value of the total electron spin $\bm I^{(e)} = \bm s_1^{(e)} + \bm s_2^{(e)}$, and its $z$-projection $I_z^{(e)}$. Here $\bm s_1^{(e)}$ and $\bm s_2^{(e)}$ are the spins of the first and the second electron. It is well known~\cite{Cortez:2002a,Glazov:2009a} that due to the electron-electron ($e$-$e$) exchange interaction, the ground state of two electrons occupying $s$- and $p$-levels in a quantum dot is a triplet state, characterized by a total momentum $I^{(e)} = 1$ and an antisymmetric combination of lateral envelopes $\psi_s(\bm \rho)$ and $\psi_{p} (\bm \rho)$:
\begin{equation}
\label{ee:triplet}
\ket{\mathrm T, m}_e = \frac{1}{\sqrt{2}} \chi_m  \left[ \psi_s(\bm \rho_1) \psi_p(\bm \rho_2) - \psi_p(\bm \rho_1) \psi_s(\bm \rho_2) \right]\:.
\end{equation}
Here $\bm \rho_1$ and $\bm \rho_2$ are the in-plane electron coordinates, $\chi_m$ is the spin-dependent part of the wave function, and $m~\equiv~I_z^{(e)} = -1,0,1$. As compared with the symmetric combination, two-electron wave function~\eqref{ee:triplet} minimizes Coulomb repulsion between the charge carriers. The two-electron singlet state is far above in energy (typically\cite{Bouet:2014a} $\sim 5$~meV) from the triplet one, and is therefore disregarded in our model.
As a result, the six states of X$^{-*}$ can be distinguished by the total angular momentum of the hole and the electron triplet $S_z (\mathrm{X}^{-*}) = J_z + I_z^{(e)} = \pm 1/2, \pm 3/2, \pm 5/2$.

In the simplest model where the trigonal effects are neglected, the electron-hole exchange interaction which defines the fine structure of the X$^{-*}$ complex at zero magnetic field is described by the following Hamiltonian
\begin{equation}
\label{eq:Xmin0}
\mathcal H_0 (\mathrm X^{-*}) = -\Delta_0 \sigma_z^{(h)} I_z^{(e)}\:,
\end{equation}
where $\Delta_0$ is the constant of the electron-hole ($e$-$h$) exchange interaction between the heavy hole and the pair of electrons in the triplet state ($\Delta_0 > 0$). 
This constant is an average
\begin{equation}
\label{eq:exch_Xminstar}
\Delta_0 = \frac12 \left( \delta_0^{ss} + \delta_0^{ps} \right)\:
\end{equation}
of the constants of the interaction between the $s$-electron and $s$-hole ($\delta_0^{ss}$) and the 
$p$-electron and $s$-hole ($\delta_0^{ps}$).\footnote{Weak cubic contributions to the exchange interaction which may mix bright and dark states at $\bm B=0$ are neglected hereafter.}
It follows from Eq.~\eqref{eq:Xmin0}, that the energy spectrum of X$^{-*}$ at zero magnetic field constitutes three Kramers degenerate levels with the angular momentum components $\pm 1/2$, $\pm 3/2$ and $\pm 5/2$, split by $\Delta_0$, see Fig.~\ref{fig:Xminstar}(d). At $\bm B=0$ only $\pm 1/2$ and $\pm 3/2$ states are optically active (see below for details) resulting in the doublet at $\bm B=0$, see Fig.~\ref{fig:Xminstar}.

The Zeeman splitting of spin-degenerate levels in the longitudinal magnetic field $\bm B  = (0,0,B_z)$ is described by the following Hamiltonian, cf. Eqs.~\eqref{electron:Z} and \eqref{eq:hh_mixing},
\begin{multline}
\label{eq:XminB}
\hat{\mathcal H}_B (\mathrm X^{-*}) = g_1 \mu_B I_z^{(e)} B_z + \\
+ \frac12 \left[ g_{h1}^{(s)} \sigma_z^{(h)} + g_{h2}^{(s)} \sigma_x^{(h)} \right] \mu_B B_z\:,
\end{multline}
where $g_1$ is the effective $g$-factor of the triplet state. Making use of the explicit form of two-electron wavefunction~\eqref{ee:triplet}, we express the triplet $g$-factor $g_1$ via $s$- and $p$-states electron $g$-factors, $g_e^{(s)}$ and $g_e^{(p)}$, as\footnote{Four $p$-shell electron states transform according to $\Gamma_3\times \Gamma_4 = \Gamma_4+\Gamma_5 + \Gamma_6$ in $C_{3v}$ point symmetry group. The magnetic field induced mixing of $\Gamma_5 + \Gamma_6$ analogous to that for heavy holes is disregarded.}
\begin{equation}
g_1 = \frac12 \left( g_e^{(s)} + g_e^{(p)} \right)\:.
\end{equation}

The total X$^{-*}$ Hamiltonian $\mathcal H(\mathrm X^{-*}) = \mathcal H_0(\mathrm X^{-*}) + \mathcal H_B(\X^{-*})$ assumes according to Eqs.~\eqref{eq:Xmin0}, \eqref{eq:XminB} the block-diagonal form
in the basis $(+5/2, -5/2, +1/2, -1/2, +3/2, -3/2)$
\begin{equation}
\label{eq:HXmin_block}
\mathcal H(\X^{-*}) = 
\left(
\begin{array}{cc}
\mathcal H_{4\times4} (\X^{-*}) & \bm 0 \\
\bm 0 & \mathcal H_{2\times2} (\X^{-*})
\end{array}
\right)\:.
\end{equation}
The blocks entering Eq.~\eqref{eq:HXmin_block} read
\begin{widetext}
\begin{equation}
\label{eq:H44}
\mathcal H_{4\times4}(\X^{-*}) = 
\left(
\begin{array}{cccc}
-\Delta_0& 0 & 0& 0\\
0& -\Delta_0& 0 & 0\\
0& 0& \Delta_0& 0 \\
0& 0 & 0& \Delta_0\\
\end{array}
\right) +
\frac12 \mu_B B_z
 \left(
\begin{array}{cccc}
2g_1 + g_{h1}^{(s)}& 0 & 0& g_{h2}^{(s)}\\
0& -2g_1 - g_{h1}^{(s)}& g_{h2}^{(s)} & 0\\
0& g_{h2}^{(s)}& -2g_1 + g_{h1}^{(s)}& 0 \\
g_{h2}^{(s)}& 0 & 0& 2g_1-g_{h1}^{(s)}\\
\end{array}
\right)\:,
\end{equation}
\end{widetext}
and
\begin{equation}
\label{eq:H22}
\mathcal H_{2\times2}(\X^{-*}) = \frac12 \mu_B B_z 
\left(
\begin{array}{cc}
g_{h1}^{(s)} & g_{h2}^{(s)} \\
g_{h2}^{(s)} & -g_{h1}^{(s)} \\
\end{array}
\right)\:.
\end{equation}
Note, that the pair $|\pm 3/2 \rangle$ is decoupled from the other states.
The structure of the block $\mathcal H_{4 \times 4}$ closely resembles the effective Hamiltonian of a neutral exciton X$^0$ in a trigonal dot.\cite{Sallen:2011a,Durnev:2013a}
The corresponding behavior of $\mathcal H_{4\times4}$ eigenenergies with the characteristic level anticrossing is shown in Fig.~\ref{fig:Xminstar}(d) by dashed and dashed-dotted curves. Since the triplet state with $I_z^{(e)} = 0$ is affected neither by the exchange interaction, nor by the magnetic field, the block $\mathcal H_{2\times2}$, Eq.~\eqref{eq:H22}, corresponds to a single heavy-hole. Its energy spectra consists of two levels with the energies~\cite{Sallen:2011a} $\eps_{\pm} = \pm \sqrt{(g_{h1}^{(s)})^2 + (g_{h2}^{(s)})^2} \mu_B B_z/2$ [see Fig.~\ref{fig:Xminstar}(d)], and the corresponding eigenstates are linear combinations of $| +3/2 \rangle$ and $| -3/2 \rangle$ states.

%\begin{figure}[htpb]
%\includegraphics[width=0.95\linewidth]{Figs/Fig4.pdf}
%\caption{\label{fig:fig4} Energy levels and schematic illustration of selection rules in $\sigma^+$ polarization for the hot negatively charged trion X$^{-*}$. The lower part of the figure shows energy levels of the $p$-electron (e$^*$) -- the state that occurs after X$^{-*}$ recombination. Dashed arrows depict the transitions that are active due to heavy-hole mixing, i.e. $g_{h2}^{(s)} \neq 0$. The parameters are $\Delta_0 = 80$~$\mu$eV, $g_1 = 0.5$, $g_e^{(p)} = 0.5$, $g_{h1}^{(s)} = 0.5$, $g_{h2}^{(s)} = 0.9$.
%}
%\end{figure}

Next we turn to the selection rules for optical transitions. Allowed optical transitions in $\sigma^+$ polarization for $B_z \geq 0$ are sketched in Fig.~\ref{fig:Xminstar}(d). The final state after optical recombination of the $s$-hole and $s$-electron is the electron in $p$-state ($e^*$), described by the Hamiltonian~\eqref{electron:Z}. In the absence of heavy-hole mixing, i.e. at $g_{h2}^{(s)} = 0$, active optical transitions in $\sigma^+$ polarization are $|+3/2 \rangle_{X^{-*}} \to | +1/2 \rangle_{e^*}$ and $| +1/2 \rangle_{X^{-*}} \to | -1/2 \rangle_{e^*}$, where the subscripts denote corresponding initial and final QD states. However, at $g_{h2}^{(s)} \neq 0$ the states $|+3/2 \rangle_{X^{-*}}$ and $|-3/2 \rangle_{X^{-*}}$, as well as $|+1/2 \rangle_{X^{-*}}$ and $|-5/2 \rangle_{X^{-*}}$ are intermixed, and therefore all the four states become optically active in $\sigma^+$ polarization. The selection rules in $\sigma^-$ polarization can be obtained by changing the sign of all spin projections. 
To summarize, the PL-spectrum of X$^{-*}$ in a given circular polarization consists of four lines in total, two of which can be referred as ``bright'' ones, i.e., active even for $g_{h2}^{(s)} = 0$, and the others as ``dark'' ones, active only because $g_{h2}^{(s)} \neq 0$.

Figure~\ref{fig:Xminstar}(c) presents the results of our calculations for the magneto-PL spectrum of X$^{-*}$ with the parameters given in the figure caption. We use the values of $g$-factors of the $s$-hole and $s$-electron that were experimentally determined from the analysis of the ground-state exciton X$^0$ using procedure explained in Refs.~\onlinecite{Sallen:2011a,Durnev:2013a}, see Tab.~\ref{tab:table1}.
The only fitting parameters are the exchange interaction constant $\Delta_0$ and $g$-factor of the electron excited state $g_e^{(p)}$. The value of $\Delta_0$ is determined from the experimental value of doublet splitting at $B_z = 0$, see Fig.~\ref{fig:Xminstar}(b). The value of $g_e^{(p)}$ is extracted from the fitting of experimental splittings of the two ``inner'' lines (corresponding to initial $\ket{\pm3/2}_{X^{-*}}$ states) in $\sigma^+$ and $\sigma^-$ polarizations. This fitting allows one to extract independently $g_e^{(p)}$ and $g_h^{(s)} \equiv \sqrt{(g_{h1}^{(s)})^2 + (g_{h2}^{(s)})^2}$. The obtained value is $g_e^{(p)} = 0.47$, which only slightly differs from $g_e^{(s)}$. The value $g_{h}^{(s)} = 0.78$, extracted from the fitting of X$^{-*}$ PL spectrum, is very close to the one presented in Tab.~\ref{tab:table1} for X$^0$ (the small difference may be related to Coulomb-induced contributions to the effective hole $g$-factor).

One can see that the model  reproduces well all the characteristic features of the X$^{-*}$ PL spectrum observed in the experiment. The crucial point for the identification of X$^{-*}$ complex is the appearance of low-energy ``dark'' transitions as the magnetic field amplitude increases. As shown above, this is due to the admixture of $|\pm1/2 \rangle_{X^{-*}}$ component in $|\mp5/2 \rangle_{X^{-*}}$ state.

\begin{table}
\caption{\label{tab:table1} $g$-factors of $s$-states and $s$-$s$ electron-hole exchange constants in the studied dots. Parameters are extracted from the analysis of the X$^0$ complex. }

\begin{ruledtabular}
\begin{tabular}{cccc}
 & QDC & QDK & QDP \\
 \hline
$g_e^{(s)}$ & 0.46 & 0.45 & 0.32 \\
$g_{h1}^{(s)}$ & 0.04 & 0.38 & 0.70\\
$g_{h2}^{(s)}$ & 0.92 & 0.76 & 0.57\\
$\delta_0^{ss}$ ($\mu$eV) & 290 & 350 & 160\\
$\delta_1^{ss}$ ($\mu$eV) & 8 & 6 & 0
\end{tabular}
\end{ruledtabular}
\end{table}

%************************************************
\subsection{Hot positively charged trion X$^{+*}$}\label{sec:Xplus_star}
%************************************************

 \begin{figure}[htpb]
\includegraphics[width=0.5\textwidth]{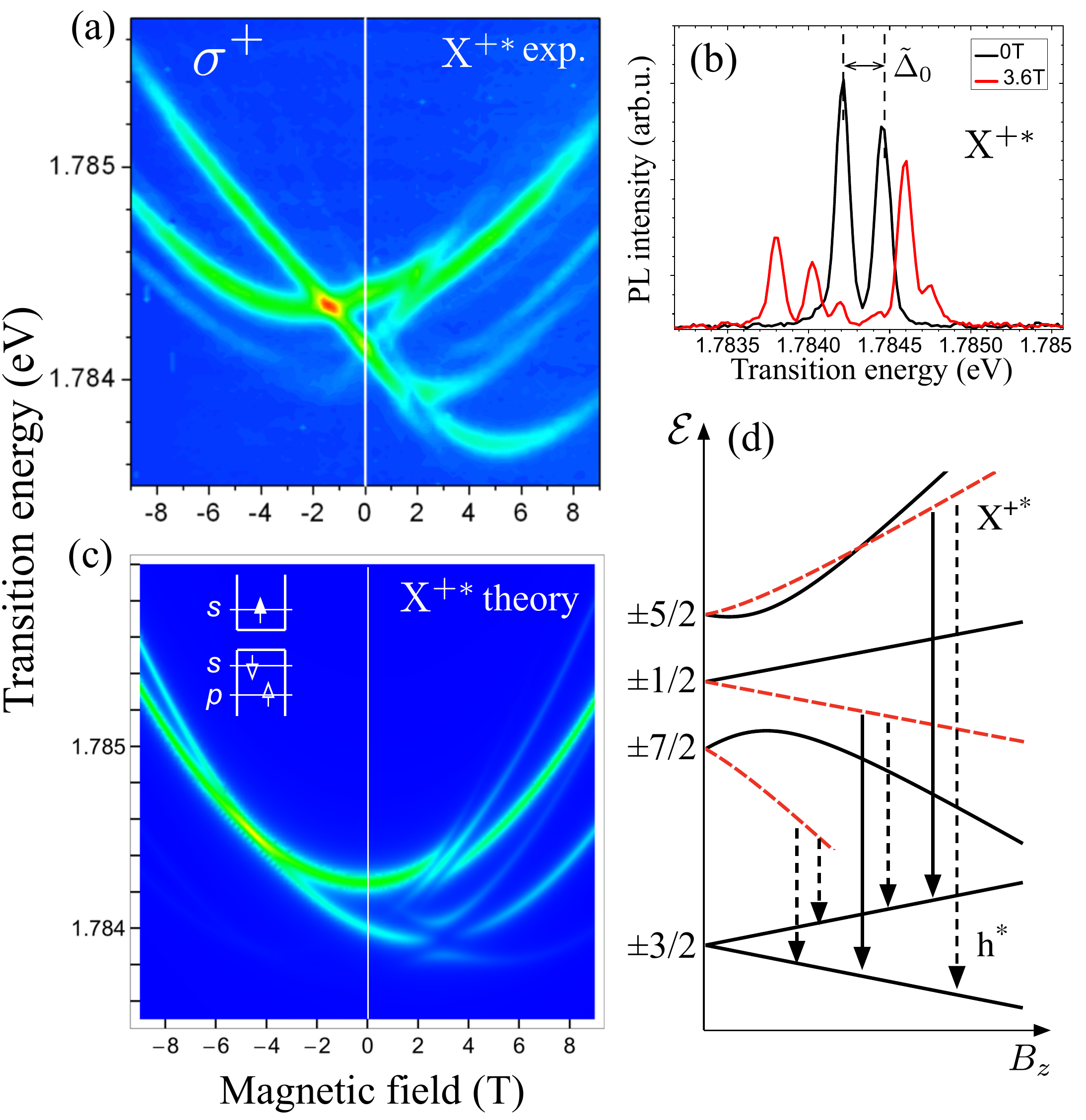}
\caption{\label{fig:Xplstar} Summary of theoretical and experimental results on magneto-PL of the X$^{+*}$ complex (QD K). (a) Measured PL pattern in $\sigma^+$ polarization and (b) measured PL spectra in $\sigma^+$ polarization at two values of magnetic field, (c) theoretical calculation of the PL pattern, (d) energy levels and schematic illustration of selection rules in $\sigma^+$ polarization. Solid and dashed lines in panel (d) show dispersions of two decoupled blocks of Hamiltonian, Eqs.~\eqref{eq:H_Xplstar} and \eqref{eq:H_Xplstar_3}. The lower part of the panel (d) shows energy levels of the $p$-hole ($h^*$) -- the state that occurs after X$^{+*}$ recombination; dashed arrows depict the transitions that become active due to heavy-hole mixing, i.e. $g_{h2}^{(s)} \neq 0$ and $g_{h2}^{(p)} \neq 0$. The parameters used in calculations are $\tilde{\Delta}_0 = 250$~$\mu$eV, $g_{h1}^{(p)} = 2.5$, $g_{h2}^{(p)} = 0.4$, $\gamma = 40$~$\mu$eV, $\alpha_d = 13$~$\mu$eV/T$^2$, where the value of $\tilde{\Delta}_0$ is directly taken from the spectrum shown in panel (b).
}
\end{figure}

We will next discuss the ``hot'' positively charged trion X$^{+*}$, see Fig.~\ref{fig:Xplstar}. This complex comprises an electron in the $s$-state and two holes, one in the $s$-shell and one in the $p$-shell state. Although the X$^{+*}$ complex is the charge conjugate of the X$^{-*}$ complex considered in the previous subsection, its fine structure in the magnetic field is very different from that of its negatively charged counterpart, as seen in Fig.~\ref{fig:Xplstar}(c) where the calculated spectrum is shown. The experimental magneto-PL spectrum of X$^{+*}$ is presented in Fig.~\ref{fig:Xplstar}(a). At zero magnetic field the PL spectrum of X$^{+*}$, as well as the spectrum of X$^{-*}$, consists of a doublet separated by 150\dots250~$\mu$eV. When a magnetic field is applied, the PL spectrum evolves into a complex pattern: the magneto PL contour plot has an asymmetric ``scissors''-like shape with two high intensity transitions for each detected circular polarization. In $\sigma^+$ polarization up to six lines are resolved at magnetic fields $B_z \gtrsim 4$ T, see Fig.~\ref{fig:Xplstar}(b). It is clearly seen that some of these lines correspond to ``dark'' transitions and appear only due to the admixture of the ``bright'' states, typical for (111) grown dots.

Similarly to the case of X$^{-*}$, we consider only two heavy-hole triplet state with antisymmetric orbital wave function, cf. Eq.~\eqref{ee:triplet}. As a result, the two-hole state can be labelled by the $z$-component of the total angular momentum of the pair, $F_z^{(h)} = -3,0,3$. Hence, the six states of X$^{+*}$ can be distinguished by the $z$-component of the total spin $S_z(\mathrm X^{+*}) = s_z^{(e)} + F_z^{(h)} = \pm 1/2, \pm 5/2, \pm 7/2$. In the absence of magnetic field the active states are $\pm 1/2$ and $\pm 5/2$ giving rise to two doublets, see below for details.
 
The effective X$^{+*}$ Hamiltonian which takes into account both $e$-$h$ exchange interaction and the Zeeman effect in the field $\bm B\parallel [111]$ has the form
\begin{multline}
\mathcal H(\X^{+*}) = -\tilde{\Delta}_0 \sigma_z^{(e)} I_z^{(h)} + \frac12 g_e^{(s)} \mu_B \sigma_z^{(e)} B_z + \\
+   \left(g_3 I_z^{(h)} + g'_3 I_x^{(h)} \right)\mu_B B_z\:,
\end{multline}
where $\sigma_z^{(e)}$ is the electron spin-$z$ Pauli matrix, and we introduced $3\times 3$ matrices of pseudospin $1$ operator $\bm I^{(h)} = (I_x^{(h)}, I_y^{(h)}, I_z^{(h)})$ as a sum of single heavy-hole pseudospin operators, $F_z^{(h)} = 3I_z^{(h)}$. The electron-hole exchange interaction constant in X$^{+*}$, $\tilde \Delta_0$, is defined by analogy with X$^{-*}$ as
\begin{equation}
\label{eq:exch_Xplstar}
\tilde{\Delta}_0 = \frac12 \left( \delta_0^{ss} + \delta_0^{sp} \right)\:,
\end{equation}
where $\delta_0^{sp}$ stands for the $s$-electron and $p$-hole exchange constant. Due to heavy-hole mixing in the longitudinal magnetic field, Eq.~\eqref{eq:hh_mixing}, the Zeeman effect for two-hole triplet is described by two $g$-factors, being related with the
$g$-factors of a single $s$- and $p$-hole as follows\footnote{In $C_{3v}$ symmetry a quadruplet of $p$-shell heavy hole states transforms according to $\Gamma_4+\Gamma_4$ representation. The magnetic field induces splitting of each of the doublets and mixing of them giving rise to two $g$-factors, $g_{h1}^{(p)}$ and $g_{h2}^{(p)}$.}
\begin{equation}
g_3 = \frac12 \left( g_{h1}^{(s)} + g_{h1}^{(p)} \right)\:,\:\:\:g'_3 = \frac12 \left( g_{h2}^{(s)} + g_{h2}^{(p)} \right)\:.
\end{equation}
The Hamiltonian $\mathcal H(\X^{+*})$ written in the basis ($+7/2, -5/2, +1/2, -7/2, +5/2, -1/2$) consists of two independent blocks
\begin{equation}
\label{eq:H_Xplstar}
\mathcal H(\X^{+*}) = 
\left(
\begin{array}{cc}
\mathcal H_{3\times3}^+ (\X^{+*}) & \bm 0 \\
\bm 0 & \mathcal H_{3\times3}^- (\X^{+*})
\end{array}
\right)\:,
\end{equation}
describing the fine structure of the hole triplet separately for the spin-up and spin-down electron and given by
\begin{widetext}
\begin{equation}
\label{eq:H_Xplstar_3}
\mathcal H_{3\times3}^\pm (\X^{+*})  = 
\left(
\begin{array}{ccc}
-\tilde{\Delta}_0 & 0 & 0\\
0 &\tilde{\Delta}_0 & 0 \\
0 & 0 & 0\\
\end{array}
\right) 
+ \frac12 \mu_B B_z \left(
\begin{array}{ccc}
\pm (g_e^{(s)} + 2g_3) & 0 & \sqrt{2}g'_3\\
0 & \pm(g_e^{(s)} - 2g_3) & \sqrt{2}g'_3 \\
\sqrt{2}g'_3 & \sqrt{2}g'_3 & \pm g_e^{(s)}
\end{array}
\right)\:.
\end{equation}
\end{widetext}
The eigenenergies of the blocks $\mathcal H_{3\times3}^+$ and $\mathcal H_{3\times3}^-$ are depicted in Fig.~\ref{fig:Xplstar}(d) by solid and dashed lines, respectively.

The final state of a QD after optical recombination of X$^{+*}$ involving $s$-shell electron and $s$-shell hole is the $p$-shell hole $(h^*)$ described by the $2\times 2$ Hamiltonian~\eqref{eq:hh_mixing} with $g$-factors $g_{h1}^{(p)}$ and $g_{h2}^{(p)}$. The selection rules in a given circular polarization are determined by the orientation of the electron spin, so that the states described by $\mathcal H_{3\times3}^+$ ($\mathcal H_{3\times3}^-$) are active in $\sigma^-$ ($\sigma^+$) polarization.
In the absence of heavy-hole mixing, i.e. at $g'_3 = 0$ and $g_{h2}^{(p)} = 0$, optical transitions active in the $\sigma^+$ polarization are $|-1/2 \rangle_{X^{+*}} \to | -3/2 \rangle_{h^*}$ and $| +5/2 \rangle_{X^{+*}} \to | +3/2 \rangle_{h^*}$. With inclusion of the heavy-hole mixing the situation becomes diverse. First of all, due to mixing of the triplet states within the blocks $\mathcal H_{3\times3}^{\pm}$, the $|\pm7/2 \rangle_{X^{+*}}$ states become optically active. Secondly, since the  $p$-holes in the final states $|\pm 3/2 \rangle_{h^*}$ are coupled, the transitions to both $p$-hole states are possible in a given circular polarization. As a result, the X$^{+*}$ magneto-PL spectrum in a given circular polarization consists of six lines in total: two ``bright'' ones and four ``dark'' ones.

Figure~\ref{fig:Xplstar}(c) presents the results of calculations of the X$^{+*}$ magneto-PL spectrum. The resemblence with the experimental results shown in Fig.~\ref{fig:Xplstar}(a) is striking. Our simulation reproduces well both the PL asymmetry with the two intense lines at $B_z < 0$ and the complicated structure at $B_z > 0$ with characteristic anticrossings of the lines and activation of the ``dark'' transitions. The total number of lines, six, is also consistent with the experimental spectrum. We note that the ``dark'' states are more pronounced in the experiment as compared with the model calculations, because their steady-state occupancies may be higher due to longer radiative decay times.

%************************************************
\subsection{Double negatively charged exciton X$^{2-}$}\label{sec:X2_min}
%************************************************

\begin{figure}[htpb]
\includegraphics[width=0.5\textwidth]{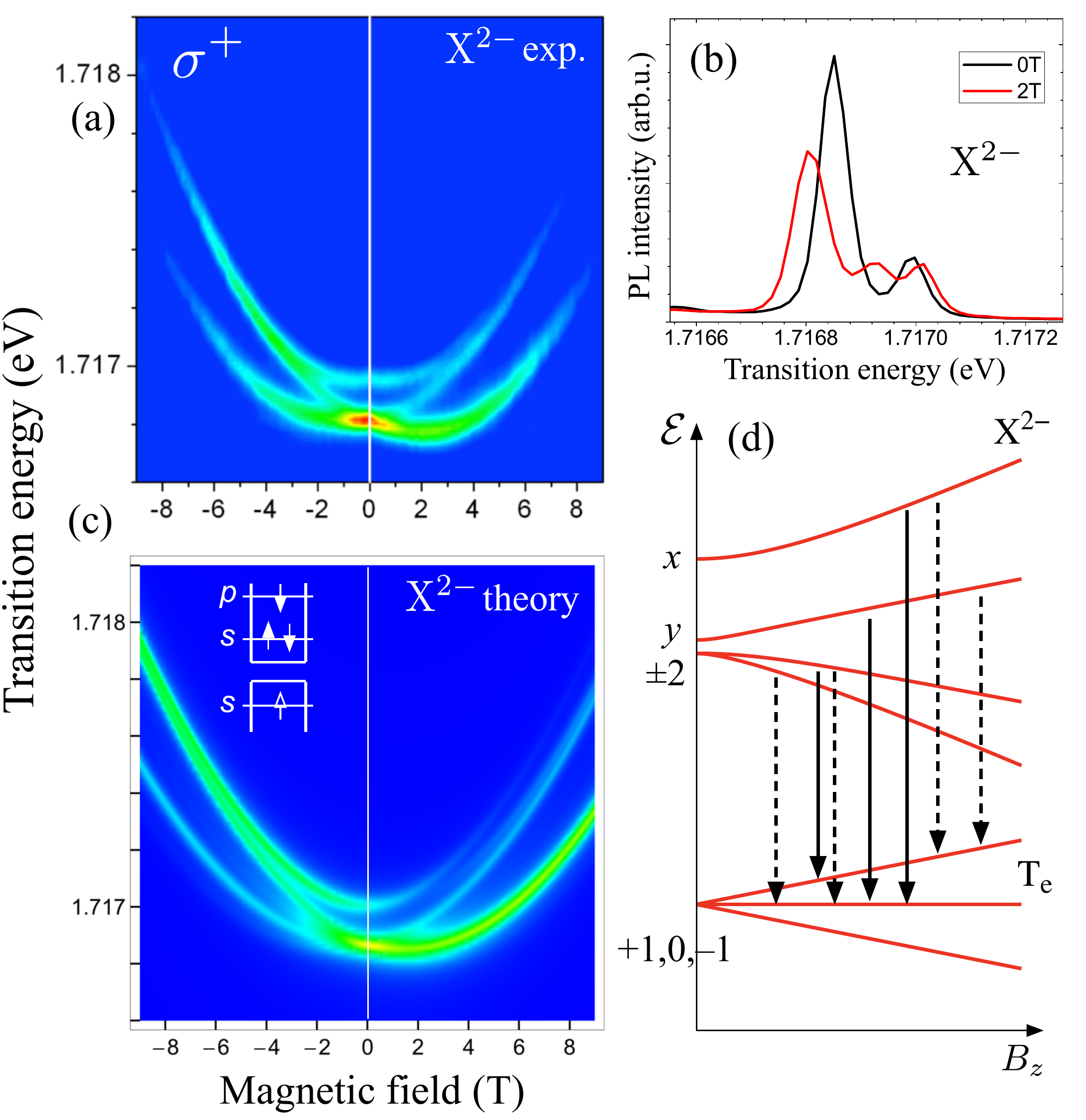}
\caption{\label{fig:X2min} Summary of theoretical and experimental results on magneto-PL of the X$^{2-}$ complex (QD P). (a) Measured PL pattern in $\sigma^+$ polarization and (b) measured PL spectra in $\sigma^+$ polarization at two values of magnetic field, (c) theoretical calculation of the PL pattern, (d) energy levels and schematic illustration of selection rules in $\sigma^+$ polarization. The lower part of the panel (d) shows energy levels of the pair of electrons ($\mathrm T_{\rm e}$) remaining after X$^{2-}$ recombination; dashed arrows depict the transitions that are active due to heavy-hole mixing, i.e. $g_{h2}^{(s)} \neq 0$. The parameters used in calculations are $\delta_0^{ps} = 70$~$\mu$eV, $\delta_1^{ps} = 140$~$\mu$eV, $g_e^{(p)} = 0.32$, $\gamma = 30$~$\mu$eV, $\alpha_d = 9$~$\mu$eV/T$^2$.
}
\end{figure}

Another important application of our theory is the analysis of the X$^{2-}$ complex, which comprises a hole and three electrons: two of them occupy the ground $s$-shell state and form the spin singlet state, and one is in the $p$-state, see the inset to Fig.~\ref{fig:X2min}(c). The final state after optical recombination of X$^{2-}$ is a pair of electrons one being in $s$-shell and the other in $p$-shell orbital state. Their spins can be either in the triplet or in the singlet state. While optical recombination to both singlet and triplet states is observed at zero magnetic field, see Fig.~\ref{fig:fig1}(c) and Ref.~\onlinecite{Bouet:2014a}, for the magneto-PL studies we focus only on the transitions to the triplet state, which are labeled as ``X$^{2-}_{\rm triplet}$'' in Fig.~\ref{fig:fig1}(c).
The measured magneto-PL spectrum of X$^{2-}$ is presented in Fig.~\ref{fig:X2min}(a). At zero magnetic field X$^{2-}$ the spectrum consists of two lines separated by 100\dots150~$\mu$eV with the lower energy transition having higher intensity. In the longitudinal magnetic field the lower energy line splits into two lines while the higher energy line does not demonstrate any splitting. As a result, at $B_z \neq 0$ there are in total three lines, see Fig.~\ref{fig:X2min}(b). 

Since the total spin of paired electrons in the singlet state is zero, the $\X^{2-}$ states are conveniently classified by the total spin of the $s$-hole and $p$-electron, which equals $\pm 1$ and $\pm 2$. Since neither exchange interaction nor magnetic field affect the singlet pair, the X$^{2-}$ complex can be viewed as the excited state of a neutral exciton X$^{0*}$ with the electron residing in the $p$-state. The only difference is that both $\pm 1$ and $\pm 2$ doublets are optically active, see below.
The effective $\X^{2-}$ Hamiltonian is given by
\begin{multline}
\mathcal{H} (\X^{2-}) = -\frac12 \delta_0^{ps} \sigma_z^{(e)} \sigma_z^{(h)} + \frac14 \delta_1^{ps} \left( \sigma_x^{(e)} \sigma_x^{(h)} + \sigma_y^{(e)} \sigma_y^{(h)} \right) \\
+ \frac12 g_e^{(p)} \mu_B \sigma_z^{(e)} B_z + \frac12 \mu_B B_z \left( g_{h1}^{(s)} \sigma_z^{(h)} + g_{h2}^{(s)} \sigma_x^{(h)} \right)\:. 
\end{multline}
Here we introduced the parameter $\delta_1^{ps}$, which gives rise to the anisotropic splitting of the $\pm 1$ doublet at $B_z = 0$. The value of  $\delta_1^{ps}$ is mainly defined by the long-range exchange interaction between the $s$-hole and $p$-electron and can be comparable or larger than $\delta_0^{ps}$ even in highly symmetric quantum dots.\cite{Glazov:2007a, PhysRevLett.98.036808} Nonzero value of $\delta_1^{ps}$ results in  formation of mixed eigenstates $\ket{x} = (\ket{-1} - \ket{+1})/\sqrt{2}$, $\ket{y} = \mathrm i( \ket{+1} +\ket{-1})/\sqrt{2}$, split in energy by $|\delta_1^{ps}|$. Here, as before, we assume that the main axes of the dot are the $x$ and $y$.

The radiative recombination of X$^{2-}$ complex involves, just like that of X$^{\pm*}$, $s$-shell electron and $s$-shell hole. As already noted we take into account only the transitions to the two-electron triplet state, described by the simple Hamiltonian  $\mathcal H (T_e) = g_1 \mu_B I_z^{(e)} B_z$. There are three ``bright'' transitions: $| x \rangle_{X^{2-}} \to | 0 \rangle_f$, $| y \rangle_{X^{2-}} \to | 0 \rangle_{T_e}$, and $| +2 \rangle_{X^{2-}} \to | +1 \rangle_{T_e}$ being active even in the absence of the hole mixing, $g_{h2}^{(s)} \equiv0$. Besides, there are four additional ``dark'' transitions active due to $g_{h2}^{(s)} \neq 0$: $| -2 \rangle_{X^{2-}} \to | 0 \rangle_{T_e}$, $|+2 \rangle_{X^{2-}} \to | 0 \rangle_f$, $|x \rangle_{X^{2-}} \to |+1 \rangle_{T_e}$, and $|y \rangle_{X^{2-}} \to | +1 \rangle_{T_e}$, see Fig.~\ref{fig:X2min}(d).

The results of our calculations of X$^{2-}$ magneto-PL spectrum are presented in Fig.~\ref{fig:X2min}(c). In our modeling we set $\delta_1^{ps} = 2 \delta_0^{ps}$, so that $\ket{y}_i$ and $\ket{\pm 2}_i$ levels are degenerate at $B_z = 0$. In that case the evolution of ``bright'' transitions with the increasing magnetic field is consistent with experiment, i.e. the lower energy lines splits into two lines, while the higher energy line remains degenerate. The low intensity ``dark'' lines are probably not resolved in the experiment. Overall, we observe rather good agreement between the theory and the experiment.

%************************************************
\section{Discussion}\label{sec:discuss}
\begin{figure*}[t]
\includegraphics[width=0.9\linewidth]{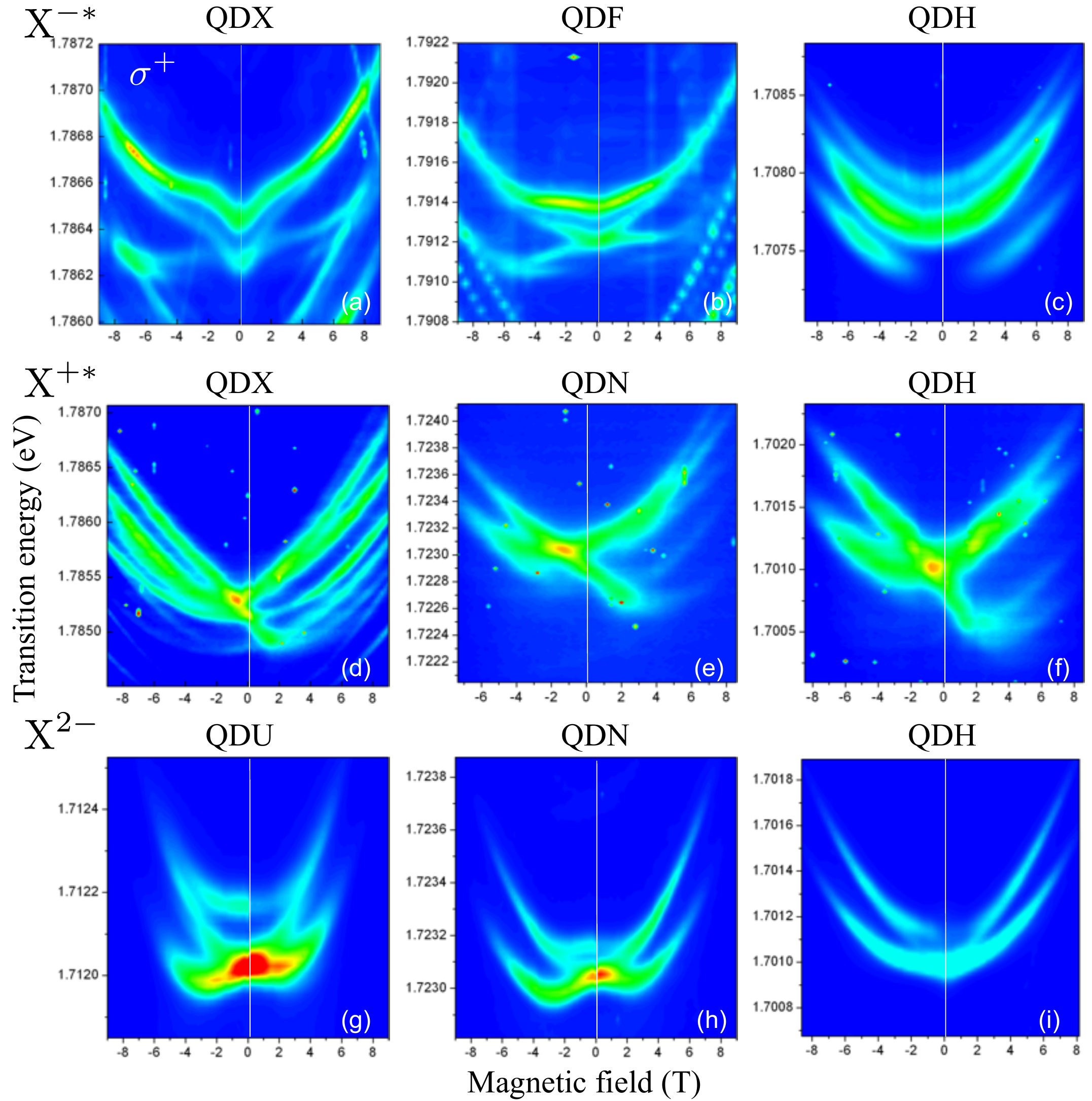}
\caption{\label{fig:more_QDs} Experimental magneto-PL patterns of three recognized excited electron-hole complexes: X$^{-*}$ (panels a-c), X$^{+*}$ (panels d-f), and X$^{2-}$ (panels g-i) presented for several dots. 
}
\end{figure*}

%************************************************
In Sec.~\ref{sec:theory_exp} we have identified transitions stemming from several electron-hole complexes in magneto-PL of single (111) QDs. Using comparison of theoretical and experimental results we were able to recognize ``hot'' trions X$^{-*}$ and X$^{+*}$ in two representative dots, QD C and QD K, respectively, and X$^{2-}$ in QD P. In fact, such an identification is not restricted to these three particular QDs, and these complexes have been observed in many studied dots in the sample. The experimental data demonstrating emission patterns of X$^{-*}$, X$^{+*}$ and X$^{2-}$ complexes in several other QDs are presented in Fig.~\ref{fig:more_QDs}. One can see that the characteristic magneto-PL spectra shown in Figs.~\ref{fig:Xminstar}, \ref{fig:Xplstar} and \ref{fig:X2min} are well reproducible from dot to dot. The asymmetry of the spectra and the appearance of the magnetic-field-induced ``dark'' lines, optically active due to the intrinsic heavy-hole mixing in these dots, serve as specific signatures of the studied complexes.

\begin{table}[t]
\caption{\label{tab:table2} The values of $p$-electron $g$-factor and electron-hole exchange constants for several dots presented in Fig.~\ref{fig:more_QDs}. The values were extracted after the fitting of PL spectra of X$^0$, X$^{-*}$ and X$^{+*}$. The values for exchange constants are given in $\mu$eV.
}
\begin{ruledtabular}
\begin{tabular}{cccccc}
 & QDC& QDX & QDN & QDU & QDF \\
 \hline
$\delta_0^{ss}$ & 290 & 280 & 260 & 290 & 240 \\
$\delta_0^{ps}$ & 90 & 110 & 160 & 190 & 170 \\ 
%$\delta_1^{sp}$ & -- & 0.38 & -- & \\
$\delta_0^{sp}$ & -- & 100 & 120 & 30 & 100 \\
$g_e^{(p)}$ & 0.47 & 0.32 & -- & -- & 0.57
\end{tabular}
\end{ruledtabular}
\end{table}

The analysis of measured PL spectra of X$^{-*}$ and X$^{+*}$ at $B_z = 0$ using Eqs.~\eqref{eq:exch_Xminstar}, \eqref{eq:exch_Xplstar} allows us to extract the values of exchange interaction constants for an $s$- and $p$-shell electrons and holes for the QDs presented in Fig.~\ref{fig:more_QDs}, see Tab.~\ref{tab:table2}. As described in Sec.~\ref{sec:Xminstar}, fitting of the Zeeman splitting between the magneto-PL lines of X$^{-*}$, allows us to extract the value of the $p$-shell electron $g$-factor $g_e^{(p)}$, see Tab.~\ref{tab:table2}. The obtained values of $g_e^{(p)}$ only slightly differ from the ones for the $s$-shell electrons. This is in agreement with model expectations, since the electron $g$-factor is controlled by the distance to the valence band,\cite{ivchenko05a} resulting in small difference between $s$- and $p$-shell $g$-factors owing to small separation between $s$- and $p$-shell states as compared to the band gap energy. By contrast, the hole $g$-factors are, to a large extent, controlled by the magneto-induced mixing of the valence band states. Hence, the values of hole $g$-factors are expected to be very sensitive to the orbital shape of the wave function.\cite{Durnev:2013a} This is in agreement with our results: To obtain a characteristic crossing of X$^{+*}$ PL lines in $\sigma^+$ polarization at $B_z < 0$ (vector $\bm B$ is antiparallel to the light propagation direction) one should use large positive values of $g_{h1}^{(p)}$ ($g_{h1}^{(p)} \approx 2 \dots 3$), while the values of $g_{h2}^{(p)}$ are comparable with those for an $s$-hole. The detailed modeling of $g$-factors for excited hole states in QDs as well as the analysis of $B_z^2$ effects, particularly, the diamagnetic shifts,\cite{Oberli:2012a} are beyond the scope of the present paper.

Finally, we note that in order to describe the excited state of the carrier in-plane motion we take into account only single, lowest in energy, $p$-orbital assuming that the dot shape is slightly anisotropic and the symmetry deviates from the perfect $C_{3v}$ point symmetry. This requirement can be relaxed for the hot positively and negatively charged trions if the spin-orbit coupling is neglected. In this case, at $\bm B=0$ the states $p_x\pm \mathrm i p_y$ are degenerate. In the presence of $\bm B\parallel z$, the degeneracy is lifted due to the orbital effect of magnetic field, $\propto \bm L \cdot \bm B$, where $\bm L$ ($L=1$) is the orbital angular momentum operator. However, since the optical transitions involve only $s$-shell electrons and holes, neither the energy position nor the intensities of the PL lines are affected by the orbital character of the wave function of the $p$-shell carrier. As a result, the magneto-PL spectra of X$^{+*}$ and X$^{-*}$ do not change if the orbital degeneracy of $p$-shell states is taken into account. By contrast, the X$^{2-}$ emission spectrum can be reproduced only if the QD anisotropy is taken into account. 
Hence, the magneto-PL of multiply charged exciton states opens the route to study weak anisotropy of such almost perfectly shaped QDs.

%************************************************
\section{Conclusions} \label{sec:concl}
%************************************************
In this paper we have presented a combined theoretical and experimental study of magneto-PL spectroscopy of the excited and multiply charged electron-hole complexes in (111)-grown GaAs quantum dots. Using the developed theory for X$^{-*}$, X$^{+*}$, and X$^{2-}$ magneto-PL, we have identified X$^{-*}$ and X$^{+*}$ trions as well as X$^{2-}$ doubly charged exciton in the experimental magneto-PL spectra of many QDs. We have revealed that the key ingredient to describe the experimentally observed behaviour of magneto-PL lines is the effect of magneto-induced mixing of $\ket{\pm 3/2}$ heavy hole states, specific to the dots of C$_{3v}$ symmetry. The characteristic magneto-PL patterns of X$^{-*}$, X$^{+*}$, and X$^{2-}$ complexes are well reproducible from dot to dot, and can be used to identify these complexes in further experiments.

\acknowledgements

We acknowledge funding from ERC Grant No. 306719, RFBR, RF president grant MD-5726.2015.2, ``Dynasty'' foundation and LIA CNRS-Ioffe RAS ILNACS. X.M. acknowledges funding from Institut Universitaire de France.

%\bibliography{chargeQD111}  

%merlin.mbs apsrev4-1.bst 2010-07-25 4.21a (PWD, AO, DPC) hacked
%Control: key (0)
%Control: author (8) initials jnrlst
%Control: editor formatted (1) identically to author
%Control: production of article title (-1) disabled
%Control: page (0) single
%Control: year (1) truncated
%Control: production of eprint (0) enabled
%

\end{document}